\documentclass[12pt]{article}
\title{\bf Systems of particles with interaction and the cluster
formation in condensed matter}

\author{{\bf Volodymyr Krasnoholovets\footnote{E-mail: krasnoh@iop.kiev.ua}
 and Bohdan Lev\footnote{E-mail:
lev@iop.kiev.ua}} \\
 {} \\
Institute of Physics, National Academy of Sciences \\
 Prospect Nauky 46,  UA-03028 Ky\"{\i}v, Ukraine}
\date{21 December 2000; 7 June 2001; 6 December 2001; \\
29 December 2002}

\begin{document}
\maketitle

\begin{abstract}
We investigate the behaviour of a system of particles with the
different character of interaction. The approach makes it possible
to describe systems of interacting particles by statistical
methods taking into account a spatial nonhomogeneous distribution
of particles, i.e. cluster formation. For these clusters are
evaluated: their size, the number of particles in a cluster, and
the temperature of phase transition to the cluster state. Three
systems are under consideration: electrons on the liquid helium
surface, particles interacting by the shielding Coulomb potential,
which are found under the influence of an elastic field (e.g.
nucleons in a nucleus), and gravitating masses with the Hubble
expansion.

\vspace{2mm}  {\bf Key words:} statistical mechanics,
interparticle interactions, clusters   \\

molecular collisions and interactions (including statistical
theories, transition state, stochastic and trajectory models,
etc.) \  --  \ 36.90.f Other special atoms, molecules, ions, and
clusters.
\end{abstract}

\section{ Introduction}
\hspace*{\parindent} Basically, statistical description of
many-particle systems concerns homogeneous states. In monograph
[1], a general method based on quantum field theory was
successfully applied to homogeneous states of condensed media. An
approach that took into account spatial nonhomogeneous states of a
system of particles was first proposed in paper [2]; the approach
was resting on the employment of the methodology of quantum field
theory as well. The most general approach to the description of a
system of particles, which included the interparticle interaction
was offered in paper [3] in which the procedure of discount of the
number of states providing for the nonhomogeneous particle
distribution was proposed. However, a correct solution of
nonlinear equations in the saddle-point is an impossible task for
the most part: the general approach does not permit one obtains
analytical solutions for a system of particles when the inverse
operator of the interaction energy does not exist. In the interim
we need to know a series of parameters which characterize possible
clusters. The major parameters are the cluster size and the
critical temperature of cluster stability. Because of that, one
can arise the question about a more simplified approach, which,
however, would allow the study of cluster formation in an
initially homogeneous system of particles.

In the present work we study a system of interacting particles
having regard to the concept proposed in papers [2,4]. Such an
approach reduces the number of variables, which determine the free
energy and introduces a peculiar "combined variable" [5] playing a
role of a typical variable that describes the thermodynamics of
the system with a spatial nonhomogeneous particle order.

For instance, as an analog of a certain combined variable one can
consider the interaction of two waves on the water surface, which
interfere in a limited range. However, after the interference the
waves depart and hence in this case the combined variable does not
bring into existence any instability of the system studied. On the
other hand, the interaction of an incident electromagnetic wave
with a polar crystal results into the polariton, a stable
formation, which is indeed described by a combined (canonical)
variable -- a mixture of the electromagnetic field and optical
phonon.

Can the combined variable give rise to the change in the
homogeneous distribution of interacting particles? We analyze such
an option considering concrete examples. Below we take a look at
the specific systems with long-range repulsion/attraction and
short-range attraction/repulsion. In particular, the approach
provides a way of estimating of the cluster size, the number of
particles in a cluster and the temperature of phase transition to
the cluster state. The approach is examined in three examples: i)
electrons on the liquid helium surface, ii) particles with the
shielding Coulomb interaction, which are found in an applied
elastic potential, and iii) gravitating masses with the Hubble
expansion.

\section{ Statistical mechanics for model systems with interaction. The
principle of states selection}

\hspace*{\parindent} Let us consider a model system of particles
with arbitrary kinds of interactions. Nevertheless, the  de
Broglie thermal wavelength $\lambda$ of a particle is taken to be
greater than the mean distant $l$ between particles. However on
the other hand, the value of $\lambda$  must be smaller than the
mean amplitude of scattering. The type of the statistics
introduces significant peculiarities to the interparticle
interaction, which, however, can be rendered by the classical
methods allowing one does away with dynamic quantum correlations.
In this case the macroscopic state of the system in question may
be specified by filling numbers $n_s$, where the symbol $s$
describes the state of the system. In other words, particles can
occupy only knots are still non-occupied. This means that the
filling numbers $n_s$ can run only two meanings: 1 (the $s$th knot
is occupied) or 0 (the $s$th knot is not occupied). Besides, the
allocation of particles in the lattice invites the type of
statistics, which set limits on the behaviour of particles, i.e.
the statistics refers particles to either bosons or fermions, see
below.

The Hamiltonian of the system of interacting particles can be
written in the form
\begin{equation}
{\cal H}(n)=\sum_s E_s n_s - \frac{1}{2} \sum_{s s^{\prime}} v_{s
s^{\prime}} n_s n_{s^{\prime}} + \frac{1}{2} \sum_{s s^{\prime}}
u_{s s^{\prime}} n_s n_{s^{\prime}}  \label{1}
\end{equation}
where $E_s$ is the additive part of the particle energy in the
$s$th state; $v_{s s^{\prime}}$  and $u_{s s^{\prime}}$ are,
respectively, the paired energy of attraction and repulsion
between particles in the states $s$ and $s^{\prime}$. Signs before
the potentials in expression (1) reflect the proper signs of the
attractive and repulsive paired energies and therefore both of the
functions $v_{s s^{\prime}}$ and $u_{s s^{\prime}}$ are positive.

Such writing of the Hamiltonian corresponds to the model in which
particles are found in sites of the three-dimensional lattice. As
will be evident from the further consideration, a specific form of
the lattice will be unneeded. However the medium will be supposed
isotropic at the transition to the continual approximation.

Note that the Hamiltonian of the model of replacement of solid
solutions [6] amounts to the Hamiltonian (1) as well. This means
that the approach proposing can describe the behaviour of defects
in solids, allows, liquids, etc. (in these cases one particle will
characterize an unfilled place and another particle that of a
filled one). The reduced solution Hamiltonian can be amounted to a
general Ising model, which takes into account an arbitrary
interaction allowing the spatial-nonhomogeneous distribution of
particles.

The partition function of the system is
\begin{equation}
Z=\sum_{\{n_{s}\}}\exp (-{\cal H}(n_s)/k_{\rm B}T)  \label{2}
\end{equation}
where the summation provides for all possible values $n_s$, i.e.,
all states of the system. Such summation can be formally
performed, if one passes to field variables according to Ref. [2].
It is known from the theory of Gauss integrals that
\begin{equation}
\exp [\frac{\nu ^2}2\sum_{ss^{\prime }}w_{ss^{\prime }}n_s{\kern
0.5pt}n_{s^{\prime }}]= {\rm Re}\int\limits_{-\infty }^\infty
D\chi \exp [\nu \sum_sn_s\chi_s-\frac 12 \sum_{ss^{\prime
}}w_{ss^{\prime }}^{-1}\chi _s\chi_{s^\prime}] \label{3}
\end{equation}
where $D\chi\equiv\Pi_s \sqrt{{\rm det} ||W_{ss^\prime}||} {\kern
1pt}\sqrt{2\pi}{\kern 1pt} d\chi_s$ implies the functional
integration with respect to the field $\chi$ and $\nu ^2=\pm 1$ in
relation to the sign of interaction ($+1$ for attractive and $-1$
for repulsion). Besides the obligatory condition
\begin{equation}
\sum_{s^{\prime \prime}}w_{ss^{\prime \prime }}^{-1}w_{s^{\prime
\prime}s^{\prime }}=\delta _{ss^\prime}  \label{4}
\end{equation}
is full-filled. In particular, condition (4) in principle could
allow one to determine the inverse operator $w_{ss^{\prime
}}^{-1}$ (in the case when it is exist) which in its turn would
permit the construction of the Green function for interacting
particles.

Let us introduce the dimensionless energies ${\tilde
w}_{ss^{\prime }}\equiv w_{ss^{\prime }}/k_{{\rm B}}T$ and $\tilde
E_s\equiv E_s/k_{{\rm B}}T$. Note that hereinafter the tilde over
a symbol means the division of the symbol by $k_{\rm B}T$.

So the partition function (2) can be represented in the form
\begin{eqnarray}
Z&=&{\rm Re}\int\limits_{-\infty }^\infty D{\kern 1pt}\phi
\int\limits_{-\infty }^\infty D{\kern 1pt}\psi \sum_{\{n_s\}}\exp
\big [-\sum_s{\tilde E}_sn_s + \sum_s(\psi _s+i\phi _s)n_s
\nonumber
\\ && -\frac 1 2\sum_{ss^{\prime }}({\tilde u}_{ss^{\prime
}}^{-1}\phi _s\phi _{s^{\prime }}+{\tilde v}_{ss^{\prime
}}^{-1}\psi _s\psi _{s^{\prime }}) \big]. \label{5}
\end{eqnarray}
We do not include the dependence on momentum of particles to the
functional integrals in the partition function (5); the sum takes
into consideration only the spatial distribution of particles and
their energy.

Now we can settle the quantity of particles in the system,
$\sum_sn_s=N$. The procedure is an equivalent to the consideration
of the canonical ensemble. For this purpose one can use the
formula known in the theory of complex variable
\begin{equation}
\frac 1{2\pi i}\oint dz\ z^{N-1-\sum_sn_s}=1.  \label{6}
\end{equation}

It makes possible to introduce the sum of the canonical ensemble
\begin{eqnarray}
Z_N &=& {\rm Re}\frac 1{2\pi i}\oint dz\int D{\kern 1pt}\phi \int
D{\kern 1pt}\psi \exp \big\{-\frac 12 \sum_{ss^{\prime }}({\tilde
u}_{ss^{\prime } }^{-1}\phi _s\phi _{s^{\prime }} + {\tilde v
}_{ss^{\prime }}^{-1}\psi _s\psi _{s^{\prime }}) \nonumber
\\ &&+(N-1)\ln z \big\} \times \sum_{\{n_s\}}\exp \big\{\sum_sn_s(\psi
_s+i\phi _s-\tilde E _s-\ln z)\big\}.  \label{7}
\end{eqnarray}
Summing over $n_s$ (note $\{n_s\}=0,\ 1$) we get
\begin{equation}
Z= {\rm Re}\frac 1{2\pi i}\int D{\kern 1pt}\phi \int D{\kern
1pt}\psi \oint d{\kern 1pt}z\ e^{{\kern 1pt} S(\phi, \psi, z)}
\label{8}
\end{equation}
where
\begin{eqnarray}
S&=&\sum_s\{-\frac 12\sum_{s^{\prime }}({\tilde u}_{ss^{\prime
}}^{-1}\phi _s\phi _{s^{\prime }}+{\tilde v}_{ss^{\prime
}}^{-1}\psi _s\psi _{s^{\prime }})+\eta \ln \big| 1+\frac \eta
ze^{-{\tilde E}_s}e^{\psi _s}\cos \phi _s \big| \}    \nonumber
\\ &&+(N-1)\ln z. \label{9}
\end{eqnarray}
Here, the function $\eta $ is equal to $\pm 1$ (Fermi or Bose
statistics), see e.g. Refs. [7-9]. This provides the means
determining the necessary procedure of the states selection that
introduces the most essential contribution to the partition
function and defines the free energy of the system. Let us set
$z=\xi +i\zeta $ and consider the action $S$ on the transit path
passing through the saddle-point with a fixed imaginable variable
${\rm Im}\ z=\zeta_0$. In this case, it stands to reason that the
action $S$, similarly to quantum field theory, is the variational
functional that depends on three variables: the fields $\phi_s$
and $\psi_s$, and the fugacity $\xi$ (here $\xi=e^{-\mu/k_{\rm
B}T}$ where $\mu$ is the chemical potential). The extremum of the
functional must be realized at solutions of the equations $\delta
S/\delta \phi_s =0$, $\delta S/\delta \psi_s =0$, and $\delta
S/\delta \xi =0$. These equations appear as the following
\begin{equation}
\sum_{s^{\prime }}{\tilde u}_{s s^{\prime }}^{-1}\phi _{s^{\prime
}}=-\frac{e^{-{\tilde E}_s}e^{{\kern 1pt}\psi _s}\sin \phi _s}{\xi
+\eta e^{-{\tilde E} }e^{{\kern 1pt}\psi _s}\cos \phi _s};
\label{10}
\end{equation}
\begin{equation}
\sum_{s^{\prime }}{\tilde v}_{s s^{\prime }}^{-1}\psi _{s^{\prime
}}=\frac{e^{-{\tilde E}_s}e^{{\kern 1pt}\psi _s}\cos \phi _s}{\xi
+\eta e^{-{\tilde E} }e^{{\kern 1pt}\psi _s}\cos \phi _s};
\label{11}
\end{equation}
\begin{equation}
\sum_{s^{\prime }}\frac{e^{-{\tilde E}_{s^{\prime
}}}e^{\psi_{s^{\prime }}}\cos \phi_{s^{\prime }}}{\xi +\eta
e^{-{\tilde E}_ {s^{\prime}}}e^{\psi_{s^{\prime }}}\cos \phi
_{s^{\prime }}}=N-1.  \label{12}
\end{equation}
Equations from (10) to (12) completely solve the problem of the
statistical description of systems with any type of interaction.
One or another of states of the system is realized in accordance
with the solution of the nonlinear equations. Among the solutions
there are solutions which correspond to the spatial nonhomogeneous
distribution of particles. So the Bose condensation is realized in
the real space because we can say that an aggregation of particles
is the Bose condensation of a sort. One of the possibilities of
cluster formation was demonstrated in Ref. [2]; the nonhomogeneous
distribution of particles which were characterized by the shielded
Coulomb interaction for both the attraction and the repulsion just
yielded stable clusters. Such type of the interaction was chosen
owing to the existence of the inverse operator of the interaction
energy in an explicit form. Nevertheless, it is a very difficult
problem to find the inverse operator among the most widespread
types of interactions. This is why, there is the necessity to
develop methods which are capable to describe the particle
distribution at any interaction. In Refs. [4,5] the approach based
on the selection of states in the form of a ''combined variable''
has been proposed. The combined variable has given us the chance
to find connections between different fields (such as $\phi_s$ and
$\psi_s$ in the present paper) along the extremal path that passes
through the saddle-point.

\section{Combined variable and the selection of states}
\hspace*{\parindent}
Let us introduce the designation
\begin{equation}
\Gamma _s=\frac{e^{-{\tilde E}_s}e^{{\kern 1pt}\psi _s}\cos \phi
_s}{\xi +\eta e^{-{ \tilde E}_s}e^{{\kern 1pt}\psi _s}\cos \phi
_s}. \label{13}
\end{equation}
The value $\Gamma_s$ includes the two field variables, $\psi_s$
and $\phi_s$, and the fugacity $\xi$. Inserting $\Gamma_s$ into
the left hand side of Eq. (12) we will see that the total sum
$\sum_s \Gamma_s$ is equal to the number of particles in the
system studied
\begin{equation}\label{14}
\sum_s \Gamma_s = N-1.
\end{equation}
This means that $\Gamma_s$ directly specifies the quantity of
particles in the $s$th state. Hence the value $\Gamma_s$ might be
considered as a characteristic of the number of particles
contained in the $s$th cluster. So it is quite possible to
consider $\Gamma_s$ as the variable of particle number and, on the
other hand, it may be called the combined variable owing to the
fact that it combines the variables $\psi_s$, $\phi_s$, and $\xi$.
Probably the introduction of the combined variable is a rough
approximation, but it admits to determine the clusterization
conditions in cases when the inverse operator of the interaction
potential can not be found.

Utilizing the variable (13) we can represent the action (9) as a
function of only two variables, $\Gamma _s$ and $\xi$, and it
makes possible to do away with the inverse operators ${\tilde
v}_{ss^{\prime }}^{-1}$ and ${\tilde u} _{ss^{\prime }}^{-1}$ in
the final expression.

If we multiply two hand sides of Eq. (11) by the function ${\tilde
v}_{ss^{\prime \prime }}$ and then summing Eq. (11) over $s$, we
obtain
\begin{equation}
\ \psi _{s^{\prime }}=\sum_s{\tilde v}_{ss^{\prime}}\Gamma_s.
\label{15}
\end{equation}
Now let us multiply the same equation (11) by $\psi_s$; then
summing it over $s$ we acquire
\begin{equation}
\sum_{ss^{\prime }}{\tilde v}_{ss^{\prime}}^{-1}\psi_{s^{\prime
}}\psi _s=\sum_{ss^{\prime}}{\tilde v}_{ss^{\prime}}\Gamma
_{s^{\prime }}\Gamma_s.  \label{16}
\end{equation}
Multiplying Eq. (10) by ${\tilde u}_{s^{\prime \prime }s}$ and
summing it over $s$ we have
\begin{equation}
\sum_{ss^{\prime }}{\tilde u}_{ss^{\prime }}^{-1}{\tilde
u}_{s^{\prime \prime }s}\phi _{s^{\prime }}=-\sum_s{\tilde
u}_{ss^{\prime \prime }} \frac{e^{-{\tilde E}_s}e^{{\kern
1pt}\psi_s}\sin \phi _s}{\xi +\eta e^{-{\tilde E}_s}e^{{\kern
1pt}\psi_s}\cos \phi_s}. \label{17}
\end{equation}
From relationship (13) one obtains
\begin{equation}
\frac 1\xi e^{-{\tilde E}_s}e^{\psi _s}\cos \phi _s=\frac{\Gamma
_s} {1-\eta \Gamma _s}.  \label{18}
\end{equation}
Let us substitute (18) into (17) taking into account Eq. (15) and
using condition (4) for the left-hand side of Eq. (17). We gain
\begin{equation}
\phi _s=-\sum_{s^{\prime}}{\tilde u}_{s^{\prime }s}{\kern
1pt}\frac 1\xi {\kern 2pt}e^{-{\tilde E}_s}{\kern 2pt}
e^{\sum_{s^{\prime \prime}}{\tilde v}_{s^{\prime \prime }s}\Gamma
_{s^{\prime \prime}}}(1-\eta \Gamma_{s^{\prime }})\sin
\phi_{s^{\prime}}. \label{19}
\end{equation}
With the formula $\sin \phi_{s^{\prime }}=\sqrt{1-\cos ^2\phi
_{s^{\prime }} }$, \  $\cos \phi_s$  can be substituted from (18)
into (19). As a result, instead of (19) we have
\begin{equation}
\phi_s=-\sum_{s^\prime}{\tilde u}_{s^\prime s}\sqrt{\frac {1}{\xi
^2} e^{-2{\tilde E}_{s^\prime}}e^{2\sum_{s^{\prime \prime
}}{\tilde v}_{s^{\prime \prime}s}\Gamma_{s^{\prime \prime
}}}(1-\eta \Gamma_{s^\prime})^2-\Gamma_{s^\prime}^2}. \label{20}
\end{equation}
Consequently using Eqs. (10) and (13) we get
   $$ \sum_{ss^\prime}{\tilde u}_{ss^\prime}^{-1}\phi_{s^\prime}\phi
_s=-\sum_s\Gamma_s \phi_s\tan \phi_s,
  $$
or in the explicit form
\begin{eqnarray}
\sum_{ss^{\prime }}{\tilde u}_{ss^{\prime }}^{-1}\phi_{s^{\prime
}}\phi_s&=&\sum_{ss^{\prime }}{\tilde u}_{ss^{\prime }}
\sqrt{\frac 1{\xi ^2}e^{-2 {\tilde E}_{s^{\prime
}}}e^{2\sum_{s^{\prime \prime }}{\tilde v}_{s^{\prime }s^{\prime
\prime }}\Gamma_{s^{\prime \prime }}}(1-\eta \Gamma_{s^{\prime
}})^2-\Gamma_{s^{\prime }}^{{\kern 1pt}2}}           \nonumber
\\ && \ \ \ \ \ \ \ \ \ \  \times \sqrt{\frac 1{\xi
^2}e^{-2{\tilde E}_s}e^{2\sum_{s^{\prime \prime }}{\tilde
v}_{ss^{\prime \prime }}\Gamma_{s^{\prime \prime }}}(1-\eta
\Gamma_s)^2-\Gamma_s^{{\kern 1pt}2}}\ ; \label{21}
\end{eqnarray}
here, allowance is made for the negative sign of the field
$\phi_s$ (see (19)).

Now using relationships (16) and (21) we can rewrite the action
(9) in the point of extremum as follows:
\begin{eqnarray}
S=&-&\frac 12\sum_{ss^{\prime }}{\tilde v}_{ss^{\prime
}}\Gamma_{s^\prime}\Gamma_s - \eta \sum_s \ln |1-\eta
\Gamma_s|+(N-1)\ln \xi - \frac 12\sum_{ss^\prime}{\tilde u}_
{ss^\prime}                       \nonumber           \\ &&\times
\sqrt{[\frac 1{\xi^2}{\kern 1pt}e^{-2{\tilde
E}_{s^\prime}}e^{2\sum_{s^{\prime \prime}}{\tilde v}_{s^\prime
s^{\prime \prime }} \Gamma_{s^{\prime \prime}}}(1-\eta
\Gamma_{s^\prime})^2-\Gamma_{s^{\prime}}^{{\kern 1pt}2}]}
                   \nonumber                           \\
&&\times  \sqrt{[\frac 1{\xi^2}{\kern 1pt}e^{-2{\tilde
E}_s}e^{2\sum_{s^{\prime \prime }}{\tilde v}_{ss^{\prime \prime
}}\Gamma_{s^{\prime \prime }}}(1-\eta \Gamma_s)^{\kern 0.5pt
2}-\Gamma_s^{{\kern 1pt}2}]}. \label{22}
\end{eqnarray}
It has been shown in Ref. [5] that the minimum of the free energy
based on expression (22) is lower than that obtained in the
framework of the mean field approximation. Indeed, let us assume
following Ref. [7] that $N_u$ and $N_v$ are numbers of particles
getting into the influence of the repulsive and attractive force
respectively. The number of particles $\Gamma_s$ in a cluster can
be put constant equals $\Gamma$. Let interaction energies in the
range of influence of these forces be change to their average
values. In this case expression (22) becomes
\begin{eqnarray}
\ &&S=-\frac 12(N-1)\big\{N_v\Gamma {\bar U}\Gamma + N_v\Gamma
{\bar V}\Gamma +2\eta \ln |1-\eta\Gamma | \big\}\ \ \ \ \
\nonumber
\\ && \ \ \ \ \ \ -\frac 12(N-1)\big[ N_u\bar U(1-\eta \Gamma
)e^{-2\bar E }e^{2N_v\bar V\Gamma ^2}\xi ^{-2}-2\ln \xi \big]
\label{23}
\end{eqnarray}
where $\bar U$ and $\bar V$ are average values of the repulsive
and attractive potential respectively. Here, in the mean field
approximation the expression in the curly brackets coincides with
the free energy. The next term, i.e. the expression in the square
brackets decreases the free energy at $\xi <1$. This inequality,
$\xi <1$, corresponds to the coherent state of a system of
interacting particles. Thus the approach that is developing can be
considered as a general methodology of the mean field theory
applying to the nonhomogeneous particle distribution.

Expression (22) is a function of only one variable $\Gamma$ and
the fugacity $\xi$ and it is applicable to any kind of the
interaction, even though the inverse operator is unknown. The most
advantage of expression (22) lies in the fact that it provides a
way of finding of such characteristics of the system as its size,
the number of particles in a cluster and the temperature of phase
transition.

Now let us pass to continual variables into the action (22).
Inasmuch as we are interested in the nonhomogeneous distribution
of particles in an indeterminate volume, let a radius $R$ be the
fitting parameter of the system studied. Assume that the density
of particles is distinguished from zero only in the cluster
volume. Then Eq. (14) can be written as
\begin{equation}
\Gamma K = N-1  \label{24}
\end{equation}
where $K$ is the number of clusters in the system and $\Gamma$
defined in expression (13) is the combined variable of the fields
$\phi_s$ and $\psi_s$, and the fugacity $\xi$ in a cluster. As is
seen from expression (24) the variable $\Gamma$ can be interpreted
as the mean quantity of particles in a cluster. Then the passage
to the continual presentation is realized by the substitution
\begin{equation}
\sum_s f_s = K {\kern 2pt} \frac 1V \int\limits_{ \rm cluster}
f(\vec r) {\kern 2pt} d {\kern 1pt}\vec r;  \label{27}
\end{equation}
here the integration extends for the volume of a cluster and
$V=\frac{4\pi}3 {\kern 1pt} g^3$ is the effective volume occupied
by one particle where $g$ is the distance between particles, i.e.,
the lattice constant.

Thus in the continual presentation  and with allowing for the
assumption (14), we can transform the action (22) to the form
\begin{eqnarray}
S&=&-\frac 1{2V^2}\int d\vec r\int d{\kern 1pt}{\vec {r^{\prime
}}}{\tilde v}(\vec r- {\vec {r^{\prime }}})\Gamma ({\vec
{r^{\prime }}})\Gamma (\vec r)\nonumber       \\ && - \frac
1{2V^2}\int d{\kern 1pt} \vec r\int d{\kern 1pt}{\vec {r^{\kern
0.5pt \prime }}}{\tilde u}(\vec r-{\vec {r^{\prime }}} )\Gamma
({\vec {r^{\prime }}})\Gamma (\vec r)       \nonumber
\\ &&\times \sqrt{e^{-2{\tilde E}(\vec r)}{\kern 1pt}e^{{\kern 1pt}2\int
{\tilde v}(\vec r-{\vec {r^{\prime }\prime }})\Gamma ({\vec
{r^{\prime \prime }}}){\kern 1pt}d{\vec {r^{\prime \prime
}}}}\frac 1{\xi ^2}(\frac 1{\Gamma \big(\vec r)}-\eta \big)^2-1}
\nonumber
\\ &&\times \sqrt{e^{-2{\tilde E}(\vec r)}{\kern 1pt}e^{{\kern 1pt}2\int
{\tilde v}(\vec r-\vec {r^{\prime \prime }})\Gamma (\vec
{r^{\prime \prime }}){\kern 1pt}d\vec {r^{\prime \prime }}}\frac
1{\xi ^2}(\frac 1{\Gamma \big(\vec r)}-\eta \big)^2-1} \nonumber
\\ && +\eta \frac 1V \int \ln |1-\eta \Gamma (\vec r)|{\kern 1pt}
d{\kern 1pt} \vec r +   (N-1) \ln \xi \label{26}
\end{eqnarray}
where the integrals are the same as in Eq. (25).

The integration is effected by the rule
\begin{eqnarray}
\frac 1V \int\limits_{\rm cluster} f(\vec r){\kern 1pt}d {\kern
1pt}\vec r &=& \frac 1{\frac{4\pi}3 g^3 } \int\limits_0^{2\pi}
d{\kern 1pt} \varphi \int\limits_0^{\pi} \sin{\kern 1pt}\theta
{\kern 1pt}d{\kern 1pt} \theta \int\limits_g^{R} f(r) {\kern
1pt}r^2 d{\kern 1pt}r \nonumber
\\ &=& \frac 1{\frac{4\pi}3 g^3 {\kern 1pt} } {\kern 2pt}4\pi
\int\limits_g^{R} f(r) {\kern 1pt}r^2 d{\kern 1pt}r; \label{27}
\end{eqnarray}
We shift here the limits of integration from 0 and $R$ to $g$ and
$R$ respectively as it allows one to eliminate the singularity of
the integrand (the same procedure is made below in expression
(29)). Thus, we can now normalize the integrals to the number of
particles $\Gamma$ in a cluster, that is,
\begin{equation}\label{28}
\frac 1V \int d \vec r = \frac 1{(4\pi/3){\kern 1pt}g^3} {\kern
3pt} 4\pi \int\limits_g^{R}r^2 d{\kern 0.7pt}r =\frac {R^3 -
g^3}{g^3} =\Gamma - 1 \cong \Gamma;
\end{equation}
the last approximation is in fact correct as we assume that
$\Gamma$ satisfies the inequality $\Gamma \gg 1$.

Relation (28) allow us to introduce the dimensionless variable
$x=r/g$ in integral (27). Thereby the rule of transformation from
summation to integration, i.e. (25), becomes
\begin{equation}
\frac 1K \sum_s f_s {\kern 2pt} =  {\kern 2pt} 3
\int\limits_1^{\Gamma^{1/3}} f(g x) {\kern 1pt}x^2 d{\kern 1pt} x
\label{29}
\end{equation}
(once again, in expressions (27) we shift the limits of
integration from 0 and $R$ to $g$ and $R$ respectively as it
allows one to eliminate the singularity of the integrand; the same
shift is made in expression (29)).

    Having integrated the action (26), we should exploit the
following relationships
\begin{eqnarray}
&& \frac 1{V^2} \int d {\kern 1pt}\vec r \int d {\kern 1pt} \vec
{r^{{\kern 1pt} \prime}} \ {\tilde v}(\vec r -\vec {r^{{\kern
1pt}\prime}}) \Gamma (\vec r) \Gamma (\vec {r^{{\kern 1pt}
\prime}}) = \frac 1{V^2} \int d {\kern 1pt}\vec r {\kern 2pt}
{\tilde v}(\vec r){\kern 1pt }\Gamma (\vec r) \cdot \int d {\kern
2pt}\vec {r^{{\kern 1pt} \prime}} {\kern 2pt} \Gamma (\vec
{r^{{\kern 1pt}\prime}}) \nonumber
\\ && = \frac 1{V^2}  \int d {\kern 1pt} \vec r \ {\tilde v}
(\vec r){\kern 1pt} \Gamma \cdot \Gamma \int d {\kern 1pt} \vec
{r^{{\kern 1pt}\prime}} = \frac {\Gamma^{\kern 1pt 2}}{V} \cdot
\int d {\kern 1pt} \vec r \ {\tilde v} (\vec r) =  3 {\kern 1pt}
\Gamma^2 \int\limits_1^{\Gamma ^{1/3}} d{\kern 1pt} x {\kern 2pt}
x^2 \tilde v(gx); \label{30}
\end{eqnarray}
\begin{equation}
\frac 1{V^2} \int d \vec r \int d{\kern 1pt} {\vec {r^{\prime
\prime}}} \ {\tilde u}(\vec r -{\vec {r^{\prime \prime}}})
\Gamma(\vec r)\Gamma({\vec {r^{{\kern 1pt}\prime \prime}}}) =
\frac 1{V^2} \int d{\kern 1pt} \vec r \ {\tilde u} (\vec r) {\kern
2pt}\Gamma^{\kern 1pt 2} = 3 {\kern 1pt} \Gamma^2
\int\limits_1^{\Gamma^{1/3}} d{\kern 1pt} x {\kern 1pt}x^2 \tilde
u(gx) {\kern 1pt} ; \label{31}
\end{equation}
With the transformations, we have used the step function: $\Gamma
(r)=\Gamma {\kern 1pt} \vartheta (\Gamma - r^3/g^3)$ where
$\vartheta ( \Gamma - r^3/g^3)=0$ if $r^3/g^3 \geq \Gamma$ and
$\vartheta(\Gamma - r^3/g^3)=1$ if $r^3/g^3 < \Gamma$. Besides,
\begin{eqnarray}
\frac 1{V^2} \int d {\kern 1pt}\vec r \int d \vec {r^{{\kern 1pt}
\prime}}{\kern 1pt} \tilde u(\vec r, {\kern 3pt} \vec {r^{{\kern
1pt} \prime}}) f(\vec r) f(\vec {r^{{\kern 1pt}\prime}}) &=& \frac
1{V^2} \int d {\kern 1pt}\vec r {\kern 2pt}\tilde u (\vec r)
f(\Gamma) \cdot \int d {\kern 1pt}\vec {r^{\kern 1pt \prime}}
{\kern 1pt} f(\Gamma) \nonumber
\\ &=& 3 f^{\kern 0.5pt 2} \int\limits_1^{\Gamma^{1/3}}
d{\kern 1pt} x {\kern 2pt} x^2
\tilde u(gx), {\kern 8pt} \label{32}
\end{eqnarray}
where $f=\sqrt{\exp[-2\tilde E + 2 \Gamma \int d {\kern 1pt} \vec
r {\kern 2pt}\tilde v (\vec r )]}$.

Now if we introduce the designations
\begin{equation}\label{33}
a = 3 \int\limits_1^{\Gamma^{1/3}} d{\kern 1pt} x {\kern 2pt} x^2
\tilde u(gx); \ \ \ \ \ \ \  b =  3 \int\limits_1^{\Gamma ^{1/3}}
d{\kern 1pt} x {\kern 2pt} x^2 \tilde v(g x),
\end{equation}
we will represent the action (26) in the following final form
\begin{equation}
S = K\cdot \Big\{ \frac 12 (a-b){\kern 1pt}\Gamma^2 -\frac 12
\frac 1{\xi^2} \Big(1 - \eta \Gamma \Big)^2 e^{-2{\tilde E} +
2b\Gamma} + \eta \ln |1-\eta \Gamma |\Big\}+ (N-1) \ln \xi .
\label{34}
\end{equation}
If we minimize the action (34) by the variable $\Gamma$, we will
be able to define the number of particles in a cluster, the
cluster size, and the temperature of phase transition that
triggers the nonhomogeneous distribution of particles. It must be
emphasized that the definition of the parameters of nonhomogeneous
formations requires only explicit forms of potentials of
interparticle interactions. Then the system itself will select the
realization which will provide for the minimum of the free energy.
Nonetheless, we assume by first the availability of clusters and
then the conditions and parameters of their existence are defined.

Studying the phase transition of clusters to the spatial
homogeneous distribution of particles, we should take into account
the behaviour of the fugacity $\xi = e^{-\mu/ k_{\rm B}T}$.
Indeed, in the classical system the chemical potential
\begin{equation}\label{35}
\mu = k_{\rm B} T \ln (\lambda_{\kern 1pt T}^3 n)
\end{equation}
 where $n$ is the particle concentration and the  de
Broglie thermal wavelength of a particle with a mass $m$
\begin{equation}\label{36}
\lambda_{\kern 1pt T} = h/\sqrt{3 m k_{\rm B} T}.
\end{equation}
 Thus in the classical case the fugacity
\begin{equation}\label{37}
 \xi = \lambda_{\kern 1pt T}^3 {\kern 1pt} n << 1.
\end{equation}
In the case of a quantum system the strong inequality does not
hold; in this case $\xi < 1 $ and $\xi$ is only a slight less than
the unit. Thereby when we investigate the action (34) searching
for clusters at the temperature $T< T_c$, we may neglect the last
term as $|\ln \xi|$ is rather small in comparison with terms which
include highest orders of $\Gamma$. However, if we consider the
action (34) scanning for the critical temperature $T=T_c$, we
should retain the last term because in this case $\xi$ trends to a
very small magnitude such that $|\ln \xi| >> 1$ and yet other
terms decrease owing to the diminution of the value of $\Gamma$.

\section{Clusters in condensed media}

\hspace*{\parindent} In this section we apply the methodology
developed above to systems of particles with long-range repulsion
(attraction) and short range attraction (repulsion).

\subsection{\it Electrons on liquid helium surface}

\hspace*{\parindent} The energy of electrostatic repulsion between
electrons can be written as $u_{ss^{\prime }}=1/4\pi \epsilon_0
\cdot Q^{{\kern 1pt} 2}/(r_s-r_{s^{\prime }})$ where $Q$ is the
effective charge of the electron in the helium film. The
attraction between electrons caused by the deformation of the
helium film can be taken as $v_{ss^{\prime }}=\frac 12\gamma_{\rm
film} (r_s-r_{s^{\prime }})^2$ where the coefficient $\gamma_{\rm
film}$ characterizes the elasticity of the helium film in respect
to the deformation caused by imbedded electrons (see, e.g. review
[10]).

In this case the parameters  $a$ and $b$ (33) of the action (34)
are transformed to the following
\begin{equation}
a = 3 {\kern 2pt} \frac {Q^{{\kern 1pt} 2} /(4\pi \epsilon_{{\kern
1pt} 0} g)}{k_{{\rm B}}T} \int\limits_1^{\Gamma^{1/3}} \frac 1x
{\kern 2pt}x^2 d{\kern 1pt} x \cong \frac {3}{8 \pi
\epsilon_{{\kern 1pt} 0}}{\kern 2pt}\frac{Q^{{\kern 1pt} 2}}{g
k_{{\rm B}} T} {\kern 2pt}\Gamma^{2/3}; \label{38}
\end{equation}
\begin{equation}\label{39}
b = 3 {\kern 2pt} \frac {\gamma_{\rm film}{\kern 1pt}g^2}{2k_{{\rm
B}}T}{\kern 2pt}\int\limits_1^{\Gamma^{1/3}} x^2 {\kern 2pt}x^2
d{\kern 1pt} x \cong \frac 3{10} {\kern 2pt} \frac{\gamma_{\rm
film} {\kern 1pt}g^2}{k_{{\rm B}}T}{\kern 2pt} \Gamma^{5/3} \ \ \
\ \ \ \ \
\end{equation}
or
\begin{equation}
a=\tilde\alpha {\kern 1pt}\Gamma ^{2/3};\ \ \ \ \ \ \
b=\tilde\beta {\kern 1pt}\Gamma ^{5/3} \ \ \ \ \ \  \ \ \ \ \ \ \
\ \ \ \label{40}
\end{equation}
where
\begin{equation}
\tilde\alpha =\frac {3}{8 \pi \epsilon_{{\kern 1pt} 0}}{\kern
2pt}\frac{Q^{{\kern 1pt} 2}}{g k_{{\rm B}} T};\ \ \ \ \ \ \ \ \
\tilde\beta = \frac 3{10} {\kern 2pt} \frac{\gamma_{\rm film}
g^2}{k_{{\rm B}}T} \label{41}
\end{equation}
and $\alpha >> \beta$ (hereafter the tilde over a symbol means the
division by $k_{\rm B} T$). Since for electrons $\eta =+1$ and the
inequality ${\tilde E}\equiv E/k_{\rm B}T\gg 1$ is legitimate
(i.e., $\exp (-2E/k_{\rm B}T)\ll 1$ where $E$ is the kinetic
energy of an electron), the action (34) can be rewritten as
\begin{equation}
 S \cong K\cdot \Big\{\frac 12 {\kern 1pt}\tilde\alpha {\kern 1pt}\Gamma
^{8/3} - \frac 12 {\kern 1pt}\tilde\beta {\kern 1pt}\Gamma
^{11/3}+ \ln \Gamma + \Gamma \ln \xi \Big\}. \label{42}
\end{equation}
The extremum of the free energy is achieved at the solution of the
equation $S^{{\kern 1pt}\prime }(\Gamma )=0$, or in the explicit
form:
\begin{equation}
\frac 43 {\kern 1pt} \tilde\alpha {\kern 1pt}\Gamma ^{5/3} - \frac
{11}{6} {\kern 1pt}\tilde\beta {\kern 1pt}\Gamma^{8/3} + \frac
1{\Gamma} + \ln \xi = 0; \label{43}
\end{equation}
Retaining leading terms in Eq. (43), i.e. highest powers to
$\Gamma$, the equation is reduced to
\begin{equation}
\frac 43 \tilde\alpha {\kern 1pt}\Gamma ^{{\kern 1pt} 5/3} - \frac
{11}{6} {\kern 1pt}\tilde\beta {\kern 1pt}\Gamma^{{\kern 1pt} 8/3}
\approx 0 \label{44}
\end{equation}
and hence the solution to it is equal to
\begin{equation}
\Gamma = \frac {8}{11}\frac {\alpha}{\beta} \equiv \frac {40}{33}
{\kern 1pt}{\kern 1pt}{\kern 1pt}   \frac {n {\kern 1pt}
Q^2}{\epsilon_0 {\kern 1pt} \gamma_{\rm film}} \label{45}
\end{equation}
where $n=(\frac {4\pi}3g^{3})^{-1}$ is the concentration of
electrons in a cluster. The sufficiency is also satisfied, i.e.,
$S^{\prime \prime }(\Gamma )>0$  with the solution (45) and
therefore it actually determines the minimum of the action (42).

The critical temperature $T_c$ is defined from the condition
$\Gamma=2$, i.e. when the temperature is so high that no more than
two particles are able to interconnect. In this case Eq. (43)
turns into
\begin{eqnarray}
k_{\rm B} T_c &&= {\kern 1pt} {\kern 1pt}  \frac {- 4{\kern 1pt}
\alpha \cdot 2^{{\kern 1pt} 8/3} + \frac {11}{2} {\kern 1pt}
\beta \cdot 2^{11/3} }{3 \ln \xi} \nonumber
\\
 && \approx - \frac {12}{\pi} {\kern 1pt}{\kern 1pt}
\Big(\frac {2{\kern 1pt}  \pi}{3} \Big)^{1/3} {\kern 2pt}{\kern
2pt} \frac {Q^2 n^{1/3}}{\epsilon_0 \ln \xi}. \label{46}
\end{eqnarray}
In the first approximation we may allow that in the critical point
the fugacity $\xi\equiv e^{-\mu/k_{\rm B}T}$ reaches its classical
meaning $\lambda_{{\kern 1pt} T}^3 {\kern 1pt} n$ (see expressions
(35) and (36)). Expanding function $\ln \xi$  in a Taylor series,
we obtain $\ln \lambda_{{\kern 1pt} T}^3 {\kern 1pt} n \approx
3(\lambda_{{\kern 1pt} T} {\kern 1pt} n^{1/3} - 1)$. Substituting
the expansion in Eq. (46) and taking into account expression (36)
for $\lambda_{{\kern 1pt} T}$, we arrive at the following equation
for $T_c$
\begin{equation}\label{47}
k_{\rm B}T_c \approx \frac {4}{\pi} {\kern 1pt}\Big(\frac {2{\kern
1pt} \pi}{3} \Big)^{1/3} {\kern 2pt}{\kern 2pt} \frac {e^2
n^{1/3}}{\epsilon_0}.
\end{equation}

 Now, let assign the concrete numerical values to parameters.
Let the charge $Q$ be the elementary charge $e$. Put $n= 2.4
\times 10^{19}$ m$^{-3}$ and $\gamma_{\rm film} = 8.4 \times
10^{-14}$ N/m. Then setting these values into expression (45) we
get the number of electrons in a cluster: $\Gamma \approx 10^8$.
For the radius of the cluster we have $R= (3{\kern 1pt}
\Gamma/4\pi {\kern 1pt} n)^{1/3} \approx 10^{-4}$ m. The critical
temperature $T_c$ estimated from expression (47) is of the order
of 100 K that is far beyond the critical temperature (4 K) of the
liquid helium film. Thus the thermodynamic condition of the liquid
helium stability sets limits on the real temperature $T_c$ of the
cluster existence.

The behaviour of electrons in the liquid helium film has been
studied by many researchers [10]. An experimental appraisal of the
mean number of electrons in a near-surface bubblon was
approximately equal to $10^8$. The radius of the bubblon was
estimated as $10^{-2}$ cm. In such a manner our qualitative
evaluation of the values $\Gamma$ and $R$ is in agreement with the
experimental data.

\subsection{\it Shielding Coulomb potential}

\hspace*{\parindent} Let particles be repulsed by the shielding
Coulomb potential $u_{ss^{\prime}}=1/4\pi\epsilon_0 \times
Q^{{\kern 1pt} 2}e^{-\kappa |r_s-r_{s^\prime}|}
\times(r_s-r_{s^{\prime}})^{-1}$, where $\kappa$ is the effective
radius of screening of nucleons in a nucleus, and attracted by the
potential $v_{ss^{\prime}}=\frac 12 \gamma (r_s-r_{s^{\prime
}})^2$ that can be created by an outside reason (so $\gamma$ is
the force constant of an off-site elastic field). In this case the
parameters $a$ and $b$ (33) of the action (34) become
\begin{eqnarray} a&=& 3 \frac{Q^{{\kern 1pt} 2}}{4\pi
\epsilon_{{\kern 1pt} 0} g k_{{\rm B}}T}
\int\limits_1^{{\Gamma}^{1/3}} \frac 1x {\kern 2pt} e^{-\kappa g
x} x^2d{\kern 1pt}x     \nonumber   \\     &=& \frac {3Q^{{\kern
1pt} 2}}{4\pi \epsilon_{{\kern 1pt} 0} \kappa^2 g^3 k_{{\rm B}}T}
\Big[ - \Big(\kappa {\kern 1pt} g {\kern 1pt} \Gamma^{1/3}+ 1\Big)
e^{-\kappa g \Gamma^{1/3}} + \big(\kappa {\kern 1pt}g + 1
\big){\kern 1pt} e^{-\kappa g}\Big].
 \label{48}
\end{eqnarray}
We will consider weight nuclei ($\Gamma >> 1$) and therefore
$\Gamma^{1/3}$ should be several times as large as the unit.
Therefore, at the first approximation we may neglect terms with
$e^{-\kappa g \Gamma^{1/3}}$ and then expression (48) becomes
\begin{equation}\label{49}
a \simeq \frac{3}{4\pi \epsilon_{{\kern 1pt} 0}} {\kern
2pt}\frac{Q^{{\kern 1pt} 2}}{\kappa^2 g^3 k_{\rm B} T}{\kern 2pt}
(\kappa g + 1){\kern 1pt} e^{-\kappa g}. \ \ \ \ \ \ \  \ \ \ \ \
\end{equation}
Now
\begin{equation}\label{50}
b= 3 {\kern 1pt}\frac {\gamma g^2}{2k_{{\rm
B}}T}\int\limits_1^{\Gamma^{1/3}} x^2 x^2 d{\kern 1pt} x \cong
\frac 3{10}{\kern 2pt}\frac{\gamma g^2}{k_{{\rm B}}T}{\kern
2pt}\Gamma^{5/3}.
\end{equation}
Let us rewrite these expressions in the form
\begin{equation}\label{51}
a = \tilde\alpha; \ \ \ \ \ \ b = \tilde\beta {\kern 1pt}
\Gamma^{5/3}
\end{equation}
where
\begin{equation}\label{52}
\tilde\alpha = \frac{3}{4\pi \epsilon_{{\kern 1pt} 0}} {\kern
2pt}\frac{Q^{{\kern 1pt} 2}}{\kappa^2 g^3 k_{\rm B} T}{\kern 2pt}
(\kappa g + 1){\kern 1pt} e^{-\kappa g}; \ \ \ \ \ \  \tilde\beta
= \frac 3{10}{\kern 2pt}\frac{\gamma g^2}{k_{{\rm B}}T}
\end{equation}
and $\alpha >> \beta$.  Let us now substitute expressions (49) and
(50) for $a$ and $b$ in the action (34). With regard for
inequality $\tilde E >> 1$ and with allowance for $\eta = + 1$ we
will get
\begin{equation}\label{53}
S\cong  K\cdot \Big\{\frac 12 {\kern 1pt}\tilde\alpha {\kern
1pt}\Gamma^{\kern 1pt 2} - \frac 12 {\kern 1pt}\tilde\beta {\kern
1pt}\Gamma^{{\kern 1pt}11/3} + \ln \Gamma + \Gamma \ln \xi\Big\}.
\end{equation}

The extremum of the action $S$ is reached with resolving the
equation $S^{\kern 1pt \prime }(\Gamma )=0$, or explicitly
\begin{equation} \tilde\alpha {\kern 1pt} \Gamma
- \frac {11}{3} \tilde\beta {\kern 1pt}\Gamma^{\kern 1pt 8/3} +
\frac 1{\Gamma} + \ln \xi = 0. \label{54}
\end{equation}

Retaining two highest order terms in Eq. (54), we obtain
\begin{eqnarray}
\Gamma  & \approx & \Big( \frac {6}{11}{\kern 1pt}
\frac{\alpha}{\beta}\Big)^{3/5}          \nonumber         \\ &=&
\frac 43 \pi^{2/5} {\kern 1pt} n {\kern 1pt}\Big\{ \frac {15}{11}
{\kern 1pt}{\kern 1pt} \frac{Q^{{\kern 1pt} 2}}{\epsilon_0 {\kern
1pt} \kappa^2 \gamma} {\kern 2pt} \Big[ \Big( \frac 3{4\pi}
\Big)^{1/3} \frac {\kappa}{n^{1/3}}+1 \Big]\Big\}^{3/5} \exp
\Big[- \frac 35 \Big( \frac 3{4\pi}\Big)^{1/3}\frac \kappa
{n^{1/3}}\Big] . \label{55}
\end{eqnarray}

The critical temperature $T_c$ of the cluster destruction is
defined from Eq. (54) if one puts $\Gamma=2$:
\begin{eqnarray}
k_{{\rm B}}T_c \approx  \frac {4{\kern 1pt} \alpha}{-\ln \xi}.
 \label{56}
\end{eqnarray}

However, having found the fugacity $\xi$, we now cannot follow the
procedure described in the previous subsection. In a nucleus, the
temperature and the thermal de Broglie wavelength are others. In
other words, we should exploit the laws of high energy physics,
i.e. that the total energy of a nucleon is linked with the
momentum of the nucleon by the relation $E=\sqrt{p^{\kern 2pt 2}
c^{\kern 1pt 2} + m^2 c^{{\kern 1pt}4}}$. Hence in this case
\begin{equation}\label{57}
\frac 32 k_{\rm B}T = c \sqrt{p^{\kern 2pt 2} + m^2c^2}
\end{equation}
and therefore
\begin{equation}\label{58}
p \simeq 3k_{\rm B}T/2{\kern 0.5pt}c.
\end{equation}
Thus the  de Broglie thermal wavelength of the nucleon in a
nucleus is determined as
\begin{equation}\label{59}
\lambda_{\kern 1pt T} = \frac {2{\kern 1pt} h}{3 {\kern
1pt}c{\kern 1pt}{\kern 1pt}k_{\rm B}T}.
\end{equation}
Expanding function $\ln \xi = \ln (\lambda_{{\kern 1pt} T}^3 n)$
in Eq. (56) in a Taylor series we obtain $\ln \xi \simeq -3$ and,
therefore, equation (53) becomes
\begin{eqnarray}\label{60}
k_{\rm B} T_c &&\approx \frac 43 {\kern 1pt} \alpha \nonumber
   \\     &&  \equiv \frac 43{\kern 1pt}
\frac {n {\kern 1pt} Q^{{\kern 1pt} 2}} {\epsilon_0 {\kern 1pt}
\kappa^2}{\kern 1pt}     \Big\{ \Big( \frac {3}{4\pi} \Big)^{1/3}
\frac {\kappa}{n^{1/3}} + 1 \Big\} {\kern 1pt}{\kern 1pt}\exp
\Big[-\Big( \frac 3{4\pi}\Big)^{1/3} \frac \kappa {n^{1/3}}\Big].
\end{eqnarray}

The results obtained in this section may account for the reasons
of the atomic nucleus stability from the microscopic viewpoint.
The shielding Coulomb potential $u=1/4\pi \epsilon_0 \times
Q^{{\kern 1pt} 2}e^{-\kappa r}/r$ is the typical nuclear (Yukawa)
potential that provides for repulsion between protons. The
potential $v=\frac 12 \gamma r^2$, which is applied to nucleons,
is ensured their mutual attraction. For example, setting $\gamma =
4 \times 10^{17}$ N/m we obtain from expression (55) $\Gamma
\approx 30$, which corresponds to the number of nucleons in a
nucleus of zinc. The critical temperature of fission of a
proton-neutron pair calculated from expression (60) results in
$T_c \sim 10^{11}$ K. Note that this value is several times of
magnitude greater than the typical temperature of a weight nucleus
(the nucleus temperature is of the order of the Coulomb repulsion
between protons in the nucleus).

\subsection{\it Gravitating masses with Hubble expansion}

\hspace*{\parindent} Gravitational physics may also be assigned to
the condensed matter [11]. The gravitational attraction between
particles, i.e. big masses, is beyond question. However the
uniform character of the Hubble expansion is in doubt. We will use
only the fact that an additional kinetic energy of particles $E_{
s s^{\prime}} = \frac 12 {\kern 1pt} m(w_s - w_{s^{\prime}})^2$ is
associated with such an expansion. The relative velocity $w_s -
w_{s^\prime}$ of particles which are found in points $s$ and
$s^\prime$, respectively, pertains to the relative distance
between the particles, since
\[
w_s - w_{s^\prime} = H (r_s - r_{s^\prime})
\]
where $H$ is the Hubble constant. In this connection, we will
consider a model system of identical gravitating masses, i.e.
stars, with the attraction $v_{s s^{\prime}} = G m^2/ (r_s -
r_{s^{\prime}})$ and the effective repulsion $u_{s s^\prime}=
\frac 12 {\kern 1pt} H^2 {\kern 1pt} m (r_s - r_{s^\prime})^2$.
For such kinds of interactions one has for the parameters $a$ and
$b$ (33):
\begin{eqnarray}
&&a= 3 {\kern 1pt}\frac {g^2 H^2 m}{2k_{\rm B}T}
\int\limits_1^{\Gamma^{1/3}} x^2 x^2 d {\kern 1pt} x \cong \frac
3{10} {\kern 1pt} \frac {g^2 H^2 m}{k_{\rm B}T} {\kern 2pt}
\Gamma^{{\kern 1pt}5/3}; \label{61}
\end{eqnarray}
\begin{equation}\label{62}
b= 3 {\kern 1pt}\frac {G m^2}{g k_{\rm
B}T}\int\limits_1^{\Gamma^{1/3}} \frac 1x {\kern 1pt} x^2 d {\kern
1pt} x \cong \frac 32 {\kern 1pt}\frac {G m^2}{g k_{\rm B}T}
{\kern 2pt}\Gamma^{{\kern 1pt} 2/3}
\end{equation}
or
\begin{equation}
a= \tilde\alpha \Gamma^{{\kern 1pt}5/3};  \ \ \ \ \ \ b=
\tilde\beta \Gamma^{{\kern 1pt}2/3} \label{63}
\end{equation}
where
\begin{equation}
\tilde\alpha =  \frac 3{10} {\kern 1pt} \frac {g^2 H^2 m}{k_{\rm
B}T}; \ \ \ \ \ \ \tilde\beta = \frac 32 {\kern 1pt}\frac {G
m^2}{g k_{\rm B}T}. \label{64}
\end{equation}
Inasmuch as gravitating masses are associated with the Bose
statistics, i.e.  $\eta = -1$, the action (34) is rewritten as
follows
\begin{equation}
S \simeq K\cdot \Big\{\frac 12 {\kern 2pt} \tilde \alpha {\kern
1pt}\Gamma^{{\kern 1pt} 11/3} - \frac 12 {\kern 2pt}\tilde \beta
{\kern 1pt}\Gamma^{{\kern 1pt} 8/3} - \ln \Gamma + \Gamma \ln \xi
\Big\} . \label{65}
\end{equation}
In expression (65) we omit the exponent term because of the
assumption that the temperature $T$ of the universe is very close
to the absolute zero (at small $T$ the exponent in expression (34)
tends to zero). Then the equation $S^{\kern 1pt \prime}(\Gamma)=0$
becomes
\begin{equation}
\frac {11}{6} {\kern 1pt}\tilde \alpha {\kern 1pt}\Gamma^{{\kern
1pt} 8/3} - \frac 43 {\kern 1pt} \tilde \beta {\kern
1pt}\Gamma^{{\kern 1pt} 5/3} - \frac 1{\Gamma} + \ln \xi = 0.
\label{66}
\end{equation}
Here, $\alpha << \beta$. Retaining highest order terms in equation
(66) we immediately obtain the solution
\begin{equation}\label{67}
\Gamma \simeq \frac {8}{11} {\kern 1pt} \frac {\beta}{\alpha}
=\frac {120 {\kern 1pt}\pi}{33}{\kern 1pt}\frac {G{\kern
1pt}m{\kern 1pt}n}{H^2}.
\end{equation}
where $n= (\frac {4 \pi}3{\kern 1pt} g^3)^{-1}$ is the
concentration of particles, i.e. stars.

The equation for the critical temperature $T_c $ that determines
the break-up of two coupled masses is obtained from Eq. (66) in
which we set $\Gamma =2$ (and the exponent term is neglected).
With the inequality $\alpha << \beta$, we get
\begin{equation}\label{68}
k_{\rm B} T_c \approx \frac {2^{2/3}}{3}{\kern 1pt} \frac
{\beta}{\ln \xi} \equiv {\kern 1pt}{\kern 1pt} \Big(\frac
{2\pi}{3}\Big)^{1/3}{\kern 1pt}{\kern 1pt}{\kern 1pt} \frac {G
{\kern 1pt} m^2 {\kern 1pt} n^{1/3}}{\ln \xi}.
\end{equation}
By definition, the fugacity $\xi = e^{-\mu/k_{\rm B}T}$. At the
transition from the Bose statistics to the classical one, the
chemical potential gradually decreases with temperature starting
from a value $\mu \sim 0$ (but still $\mu < 0$) to $\mu << 0$. So,
the function $\ln \xi$ in Eq. (68) is strongly negative: $\ln \xi
= - |\mu|/k_{\rm B}T$. This signifies that the critical
temperature in the system of gravitating masses is absent. In
other words, in the universe, masses trends to the aggregation at
any conditions.

Thus the results obtained demonstrate that the availability of the
Hubble expansion is the deciding factor in the formation of
galaxies in the universe (note that without $H$ masses in the
universe will trend merely to the total mergence and any spatial
distribution will not be observed). Actually, expression (67)
shows that Hubble's constant, $H \approx 50$ km/(s$\cdot$Mpsc) =
$1.6 \times 10^{-18}$ s$^{-1}$, is one of the main parameters that
determines the steady state of masses in a cluster, i.e., galaxy.
So the greater the mean mass $m$, the greater the number $\Gamma$
of stars in a galaxy.

\section{Conclusion}
\hspace*{\parindent} In the present paper the statistical
description of a system of interacting particles, which allows the
study of their spatial nonhomogeneous distribution, has been
proposed. One of the most important properties of our model is the
presentation of the total potential of interacting particles in
the form of the paired energy of attraction and repulsion. We have
started from the Hamiltonian that describes interacting particles
located in knots of the three-dimensional lattice and applied the
quantum field theory methodology based on the functional integrals
for the description of the partition function of systems studied.

We have shown that the solutions of equations characterizing the
behaviour of the field variables yield the full resolving the
initial statistical task. The solutions may correspond to both the
homogeneous and nonhomogeneous distributions of particles and only
the character of interparticle interaction and the temperature
define the kind of the solution.

The free energy represented through field variables in the
saddle-point has allowed us to investigate special conditions,
which provide for the formation of clusters in a system with the
initially homogeneous particle distribution. The conditions are
answerable also for the size of clusters and the temperature of
phase transition to the cluster state.

The approach proposed does not need fitting parameters: the
cluster size and the temperature of phase transition to the
cluster state are completely determined by the value and character
of paired potentials in the systems of interacting particles.
Besides, the introduction of a peculiar type of the combined
variable has simplified the investigation procedure. Specifically,
the procedure has provided a way for the obtaining the necessity
criterion of the transition of a system of separate particles into
the spatial nonhomogeneous state.

The authors are very thankful to the referees for the concrete
constructive remarks that allowed a significant improvements of
the contents of the paper.

\end{document}